\documentclass[twocolumn,preprintnumbers,amsmath,amssymb,superscriptaddress,nofootinbib,longbibliography,prb]{revtex4-2}

\usepackage{graphicx}
\usepackage{bm}
\usepackage{dsfont}
\usepackage[usenames,dvipsnames]{xcolor}
\usepackage{pstricks}
\usepackage[tight]{subfigure}
\usepackage{verbatim}
\usepackage{units}
\usepackage{multirow}
\usepackage{mathrsfs}
\usepackage{leftidx}
\usepackage{xspace}
\usepackage{diffcoeff}  
\usepackage{bbm}
\usepackage{amsfonts,amssymb,amsmath}
\usepackage{mathtools}

\usepackage{array} 
\usepackage{tabularx}
\usepackage[normalem]{ulem}
\usepackage{bbold}
\usepackage{pifont}
\usepackage{hyperref}
\hypersetup{colorlinks=true,linktoc=all,linkcolor=blue,breaklinks=true,citecolor=blue,urlcolor=blue}

\usepackage[utf8]{inputenc}
\usepackage[english]{babel}

\addto\captionsenglish{}

\AtBeginDocument{%
    \newwrite\bibnotes
    \def\bibnotesext{Notes.bib}
    \immediate\openout\bibnotes=\jobname\bibnotesext
    \immediate\write\bibnotes{@CONTROL{REVTEX42Control}}
    \immediate\write\bibnotes{@CONTROL{%
    apsrev42Control,author="08",editor="1",pages="1",title="0",year="1"}}
     \if@filesw
     \immediate\write\@auxout{\string\citation{apsrev42Control}}%
    \fi
}%

\newcommand*{\e}{\text{e}} 
\newcommand*{\Tr}{\text{Tr}}

\newcommand{\beq}{\begin{equation}}
\newcommand{\eeq}{\end{equation}}

\newcommand*{\dQ}{\dot{Q}}
\newcommand*{\tta}{{1}}
\newcommand*{\ttb}{{2}}

\DeclarePairedDelimiterX\braket[2]{\langle}{\rangle}{#1\,\delimsize\vert\,\mathopen{}#2}
\DeclarePairedDelimiterX\ketbra[2]{\lvert}{\rvert}{#1\,\delimsize\rangle\mathopen{}\delimsize\langle\,\mathopen{}#2}

\DeclarePairedDelimiterX\Braket[2]{(}{)}{#1\,\delimsize\vert\,\mathopen{}#2}
\DeclarePairedDelimiterX\Ketbra[2]{\lvert}{\rvert}{#1\,\delimsize)\mathopen{}\delimsize(\,\mathopen{}#2}

\newcommand{\kB}{k_\text{B}}
\newcommand{\kBT}{k_\text{B}T}

\begin{document}

\title{
\vspace*{-1.25cm}
\textnormal{{\small PHYSICAL REVIEW A {\bf 110}, 022210 (2024)}}\\
\vspace*{-0.2cm}
\rule[0.1cm]{18cm}{0.02cm}\\
\vspace*{0.285cm}
All-thermal reversal of heat currents using qutrits
}
\author{Irene Ada Picatoste}
\affiliation{Institute of Physics, University of Freiburg, Hermann-Herder-Stra\ss e 3, D-79104 Freiburg, Germany}
\affiliation{Departamento de F\'isica Te\'orica de la Materia Condensada, Universidad Aut\'onoma de Madrid, 28049 Madrid, Spain\looseness=-1}
\author{Rafael S\'anchez}
\affiliation{Departamento de F\'isica Te\'orica de la Materia Condensada, Universidad Aut\'onoma de Madrid, 28049 Madrid, Spain\looseness=-1}
\affiliation{Condensed Matter Physics Center (IFIMAC), Universidad Aut\'onoma de Madrid, 28049 Madrid, Spain\looseness=-1}
\affiliation{Instituto Nicol\'as Cabrera, Universidad Aut\'onoma de Madrid, 28049 Madrid, Spain\looseness=-1}
\date{\today}

\begin{abstract}
Few-level systems coupled to thermal baths provide useful models for quantum thermodynamics and to understand the role of heat currents in quantum information settings. Useful operations such as cooling or thermal masers have been proposed in autonomous three-level systems. In this work, we propose the coherent coupling of two qutrits as a simultaneous refrigerator and heat pump of two reservoirs forming a system. This occurs thanks to the coupling to two other reservoirs which are out of equilibrium but do not inject heat in the system. We explore the thermodynamic performance of such operation and discuss whether it can be distinguished from the action of a Maxwell demon via measurements of current fluctuations limited to the working substance. 
\end{abstract}

\maketitle

\section{Introduction}

The  coupling of quantum systems to thermal baths has activated the field of quantum thermodynamics in the last years~\cite{binder:2018,Strasberg2022}, favored by a considerable experimental advance in the control of heat flows at the nanoscale~\cite{giazotto:2006,courtois_electronic_2014,pekola_colloquium_2021}. Seminal proposals six decades ago~\cite{scovil_three_1959,geusic_three_1959,geusic:1967} already considered three-level systems as constituents of minimal quantum thermodynamic machines~\cite{alicki_quantum_1979,kosloff_quantum_1984,geva_three_1994,geva_quantum_1996,boukobza_three_2007}. Recently few-level systems (qubits, qutrits...) have been proposed as autonomous heat engines
and refrigerators under the influence of two or more heat or work reservoirs~\cite{palao:2001,linden:2010prl,levy:2012,correa:2013}, see Refs.~\cite{kosloff:2014,myers_quantum_2022,liliana_review,cangemi_quantum_2023} for recent reviews. For these operations, the various system-bath couplings need to be either spatially separated (with different baths acting on different qubits) or appropriately filtered, an experimental difficulty that has been overcome recently using superconducting circuits~\cite{ronzani_tunable_2018,senior_heat_2020,gubaydullin_2022,lu_steady_2022,aamir_engineering_2022,aamir_thermally_2023,sundelin_quantum_2024,upadhyay_microwave_2024} or laser-emulated reservoirs in trapped atoms~\cite{rossnagel_single_2016}, ions~\cite{maslennikov_quantum_2019,vonlindenfels_spin_2019}, nitrogen vacancies in diamond~\cite{klatzow_experimental_2019}, nuclear spins~\cite{peterson_experimental_2019} or photons~\cite{passos_optical_2019}. From a practical point of view, quantum thermodynamic machines open possibilities to the onchip manipulation of heat flows in quantum processors, e.g. in the form of thermal transistors~\cite{ojanen_mesoscopic_2008}, rectifiers~\cite{segal_spin_2005,ojanen_selection_2009,guo_quantum_2018,kargi_quantum_2019,xu_heat_2021,iorio_photonic_2021,diaz_qutrit_2021}, switches~\cite{karimi_coupled_2017}, or transducers~\cite{thomas_thermally_2020,cao_quantum_2022}. 
Electronic analogues have also been implemented~\cite{thierschmann:2015,thierschmann_thermal_2015,josefsson_quantum_2018,dorsch:2020} with the charge occupation defining the few-state system.

The state of a qubit is clearly of information nature. 
The connection between thermodynamics and information~\cite{szilard:1929,landauer:1961,bennett:1982} unveiled by the Maxwell demon is explicit in quantum and mesoscopic setups, where one has direct access to the microscopic state of single particles, see detailed discussions for mesoscopic electronic transport in Refs.~\cite{datta:2008,maruyama:2009,parrondo:2015,pekola:2015,rob_demonrev}. The information of the qubit state can be used, via appropriate measurement and feedback mechanisms, to manipulate the thermodynamic flow~\cite{lloyd_quantum_1997,kieu_second_2004,quan_maxwells_2006,quan:2007,schaller:2011,averin_maxwell_2011,vidrighin:2016,camati_experimental_2016,elouard_extracting_2017,cottet:2017,chida:2017,schaller:2018,masuyama_information_2018,naghiloo_information_2018,annbyandersson_maxwell_2020,hernandezgomez_nonthermal_2021}. In particular, the controlled exchange coupling of two qubits can define a Maxwell demon refrigerator~\cite{lloyd_quantum_1997,quan_maxwells_2006} whose protocol can be interpreted in terms of analogue quantum heat engines~\cite{quan:2007}.
Information-based engines can also be made autonomous, allowing for a full thermodynamic interpretation in terms of measurable heat and particle currents~\cite{hotspots,sanchez_detection_2012,strasberg:2013,koski:2014,thierschmann:2015,koski:2015,horowitz:2014prx,borrelli_fluctuation_2015,ptaszynski:2018,strasberg:2018,erdman_absorption_2018,najerasantos_autonomous_2020,mayrhofer:2021,bhandari_minimal_2021,poulsen_quantum_2022}. 
Ideally one asks the autonomous demon to violate the second law while simultaneously respecting the first law in some part of the device where measurements are carried (hereforth called simply the system). For this aim, the demon needs to be comprised of at least two reservoirs so it holds a nonequilibrium situation~\cite{rob_demonrev}. Proposals so far mostly focus on electronic configurations~\cite{whitney:2016,ndemon,ptaszynski_thermodynamics_2019,sanchez:2019,deghi_entropy_2020,fatemeh,ciliberto_autonomous_2020,lu_unconventional_2021,freitas_2021,ryu_beating_2022,freitas_maxwell_2022}. 

\begin{figure}[b]
\includegraphics[width=\linewidth]{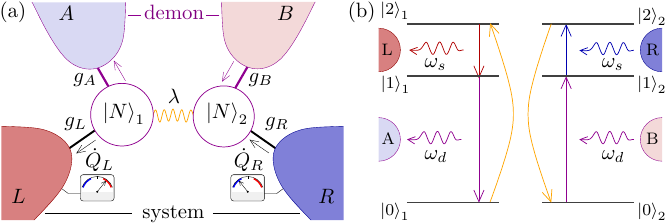}
\caption{\label{fig:scheme}
  Scheme of our device. (a) Two qutrits, $\alpha=\tta,\ttb$, coupled via an exchange interaction $\lambda$ and connected to thermal baths that form the measured system (L and R) and the demon (A and B). Couplings $g_l$ are filtered at frequencies ($\omega_s$ or $\omega_d$) determining transport in the system or the demon baths. (b) Transitions between the different states $|N\rangle_\alpha$ of each qutrit are affected by a different bath. 
  The sketched cycle achieves the transport of a quantum of heat $\hbar\omega_s$ from the cold (R) to the hot (L) system reservoirs enabled by $\hbar\omega_d$ flowing through the demon terminals (from B to A).
}
\end{figure}

In this work, we propose an autonomous demon in an all-thermal setup, in the sense that work sources are absent and transport is purely due to heat currents responding to temperature differences. We consider two qutrits ({\tta} and {\ttb}) and four photonic reservoirs, all treated on equal footing, see Fig.~\ref{fig:scheme}(a). Transport is measured in reservoirs L and R  (the system) at temperatures $T_L$ and $T_R$. The other baths (the demon reservoirs, A and B) are used to induce a  nonequilibrium situation by maintaining a temperature difference $T_B\ge T_A$. 
Each qutrit exchanges photons with one system and one demon reservoir: {\tta} with A and L, {\ttb} with B and R. The connection that enables transport between the two qutrits is via swap interactions, see Fig.~\ref{fig:scheme}(b) for a representative sequence. 

The demonic action in this case corresponds to a reversal of heat currents in the measured system under a finite temperature difference (say $T_L>T_R$), without heat being injected from the other baths (A and B) forming an environment, e.g., when heat flows out of the colder R ($\dot{Q}_R<0$) and into the hotter L ($\dot{Q}_L>0$) with $\dot{Q}_L+\dot{Q}_R=0$. It hence simultaneously enables cooling and pumping into the system. This can even happen when L and R are the hottest and coldest reservoirs, while the demon baths are only {\it warm} (we will refer to this situation as the warm demon operation). 
Of course, a proper (according to the laws) thermodynamic behaviour is recovered once one has access to the dynamics (currents) of the whole system. 

Different kinds of autonomous demons have been identified. Bipartite systems allow for mechanisms  with a clear interpretation in terms of measurement and feedback protocols based on the interpartition interactions~\cite{whitney:2016,ptaszynski_thermodynamics_2019,sanchez:2019,freitas_maxwell_2022}.
However the notion of information is not always obvious: systems coupled to nonequilibrium environments (also known as N-demons) can be tuned to achieve a demonic effect in particular configurations~\cite{ndemon,deghi_entropy_2020,fatemeh,ciliberto_autonomous_2020,lu_unconventional_2021} (i.e., if instead of the detailed knowledge of the single particle states, the demon has a global knowledge of the system~\cite{kieu_second_2004}), and by allowing fluctuating deviations of the demon conditions, (i.e., the system and the demon only exchange noise~\cite{acciai_constraints_2023}). 
Our configuration is of none of these kinds: despite using information states, it is not bipartite and has no clear interpretation of a memory; unlike N-demons there is a spatial separation of the demon sources and does not require fine tuning. 

The question of how to classify these demonic operations based on the limited information accessible in the system has attracted some interest~\cite{freitas_2021,rob_demonrev}. In other words: if an observer who can only measure the currents in two reservoirs  detects a demonic action, how can they learn about the type of demon? A criterion for a device to behave as a so-called {\it strict} Maxwell demon has been proposed based on the presence of an internal current in the device being reversed under the action of the demon~\cite{freitas_2021}. Typically this current is either not accessible in a mesoscopic device, as one measures currents in the reservoirs, or cannot be measured without affecting the nonlocal transport~\cite{buttiker_four_1986,butcher:1990}, see a pedagogical discussion in Ref.~\cite{girvin_book}. To overcome this limitation, the perfect crosscorrelation of the two system currents is suggested as a signature of such demonic process. Another desired property  of a strict demon is that the conservation of heat in the system occurs not only on average but even at the level of the fluctuations i.e., the separation of system and demon currents always holds in the stationary regime (not relying on a particular set of parameters), according to Ref.~\cite{freitas_2021}.
The spatial separation of the system and demon terminals in our model allows us to define an interface for the internal current at the coupling between the qutrits and explore its properties in connection with the reversal of the system currents.
As a clear difference with previous electronic proposals based on matter/charge currents, in our device this internal current is not continuous, in the sense that particles are being injected from some reservoirs and absorbed by others after going through the system. 
The model and relevant processes are discussed in detail in Sec.~\ref{sec:model}.

We are hence interested in the properties of heat currents and their correlations. 
To compute them, we use a full counting statistics approach~\cite{levitov_charge_1993,*levitov_electron_1996,nazarov_quantum_2003},
a  method that has been applied to electronic~\cite{bagrets_2003,christian,franz,andrieux:2009}, bosonic~\cite{segal_current_2018,friedman_quantum_2018} or mixed systems~\cite{sanchez:2007,*sanchez:2008,krause:2011} described by master equations. 
We extend a recursive method~\cite{franz}, so far restricted to the autocorrelations in charge conductors, to multi-mode photon transport through a few-level system coupled to multiple thermal baths. The method is described in Sec.~\ref{sec:transport}. On top of giving information about the system dynamics, currents and fluctuations can be used to characterize the thermodynamic performance in terms of useful power, efficiency and noise. For a multiple-reservoir and multitask performance like ours, we need to use generalized efficiencies in terms of free-energies~\cite{manzano:2020,fatemeh,tesser_thermodynamic_2023}. For the noise, we compare to the thermodynamic uncertainty relation (TUR)~\cite{barato_tur_2015,pietzonka_universal_2018,horowitz_thermodynamic_2020} as defined for classical Markovian dynamics, as discussed in Sec.~\ref{sec:performance}.

Experimentally, the exchange coupling of two qutrits can be achieved in different configurations, in particular including superconducting implementations~\cite{liu_controllable_2006,liu_controllable_2014}, see Ref.~\cite{cattaneo_engineering_2021} for a review. In the various physical implementations (e.g., superconducting circuits, atoms, quantum dots) the coupling mechanism can be very different (inductive~\cite{menke_demonstration_2022} or capacitive~\cite{aamir_engineering_2022,luo_experimental_2023} coupling, or via spin~\cite{fu_experimental_2022} or electronic exchange~\cite{shinkai_correlated_2009} interactions, among others). Coupling via a mediator~\cite{crescente_enhancing_2022} can be used to further increase the spatial separation of the circuit components and introduces further control on the frequency of the exchange~\cite{ding_high_2023}. For the sake of simplicity and generality, we will however keep our description at a phenomenological level. 
Superconducting resonators can be used to achieve the qutrit-reservoir coupling~\cite{lu_steady_2022,aamir_engineering_2022}, which has the additional advantage to facilitate the introduction of temperature differences.

The properties of heat transport through the system are discussed for different configurations of the couplings in Secs.~\ref{sec:demon}, \ref{sec:perfectfilters} and \ref{sec:lessideal}, with other possible realizations presented in Sec.~\ref{sec:otherdemons}. Conclusions are presented in Sec.~\ref{sec:conclusions}.

\section{Description of the model}
\label{sec:model}

The system we consider is composed of two thermal baths, L and R, at temperatures $T_{L/R}=T\pm\Delta T_s/2$. Their coupling is mediated by two identical coupled qutrits $q$=1,2, each of them with states $|i\rangle_q$, $i$=0,1,2, and energies $E_{iq}$, as represented in Fig.~\ref{fig:scheme}(b). Each qutrit is connected to one additional thermal bath, A or B, as depicted in Fig.~\ref{fig:scheme}. We assume the qutrits to be weakly coupled to all baths. Baths A and B are out of equilibrium with respect to each other for holding a temperature difference $\Delta T_d\geq0$ applied symmetrically: $T_{A/B}=T\mp\Delta T_d/2$. 
At $\Delta T_s=\Delta T_d=0$ the device is in equilibrium. The warm demon will operate when $\Delta T_d<|\Delta T_s|$.

\begin{figure}[t]
\includegraphics[width=0.8\linewidth]{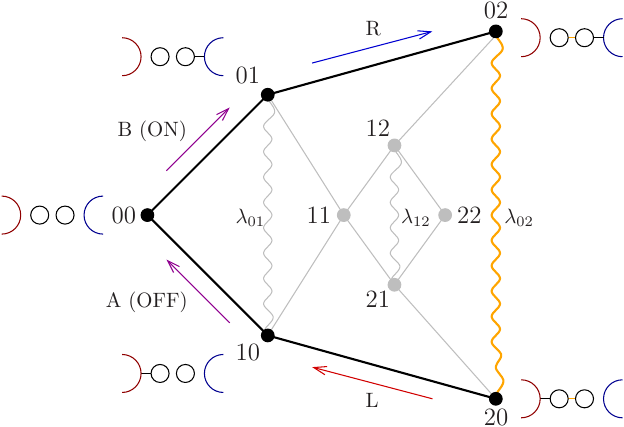}
\caption{\label{fig:cycle}
  The noisy switch demon. Solid lines indicate the transitions  between the different states of the qutrits. Black lines emphasize the basic-cycle transitions with a single excited qutrit. In the upper branch (between $|00\rangle$ and $|02\rangle$, bath L is effectively uncoupled from the qutrits (see text). Correspondingly for the lower branch (between $|20\rangle$ and $|00\rangle$) and bath R. Clockwise circulation (as marked by the arrows labeled with the reservoir involved in each transition) carries a photon from the cold bath R to the hot L. The orange wavy line indicates the primary coherent swap transition between states $|02\rangle$ and $|20\rangle$ with coupling $\lambda_{02}$. }
\end{figure}

The whole device is modeled with the Hamiltonian $\hat H_S=\hat H_0+\hat H_{1-2}$, being 
\begin{align}
\label{eq:hqutrit}
\hat H_0&=
\sum_{i=0}^2(E_{i\tta}|i\rangle_{\tta}{}_{\tta}\langle i|\otimes\mathbb{1}_{\ttb}+E_{i\ttb}\mathbb{1}_{\tta}\otimes|i\rangle_{\ttb}{}_{\ttb}\langle i|), \\
\label{eq:hab}
\hat H_{\tta-\ttb}&=\lambda_{02}|20\rangle\langle02|+\lambda_{01}|10\rangle\langle01|+\lambda_{12}|21\rangle\langle12|+{\rm h.c.}
\end{align}
the qutrit Hamiltonians and their interaction, respectively.
Here, we introduce the notation $|ij\rangle\equiv|i\rangle_1\otimes|j\rangle_2$. 
The coupling to the reservoirs is given by
\begin{equation}
\hat H_{{\tta\ttb}{-}res}=\sum_{l,q,j,k}g_{lq}(\hat a_l^{}+\hat a_l^{\dagger})\hat Y_{q,jk},
\end{equation}
where $\hat a_l^{}$ annihilates a photon in reservoir $l$=A,B,L,R, and $Y_{q,jk}=|j{\rangle_{qq}\langle} k|$ are jump operators in qutrit $q$. They induce the transitions between the different states as represented in Fig.~\ref{fig:cycle} by the exchange of photons, which leads to the heat currents $\dot{Q}_l$ out of reservoirs $l$. The couplings $g_{lq}$ are also assumed to be narrow functions of the frequency (e.g. by being mediated by filters): the system reservoirs L and R are filtered at $\hbar\omega_s=E_{2q}-E_{1q}$, and the demon ones, A and B, at $\hbar\omega_d=E_{1q}-E_{0q}$. This way, each reservoir induces a single transition in the qutrit it is coupled to: A(B) between $|0\rangle_{1(2)}$ and $|1\rangle_{1(2)}$, and L(R) between $|1\rangle_{1(2)}$ and $|2\rangle_{1(2)}$. We will consider this perfectly filtered configuration throughout the manuscript, and relax this assumption in Sec.~\ref{sec:leakage}. As we assume local couplings~\cite{landi_nonequilibrium_2021}, we have $g_{A\ttb}=g_{L\ttb}=g_{B\tta}=g_{R\tta}=0$. In the following, we consider symmetric couplings for the remaining ones and drop the qutrit index, $g_{l,q}=g_l$. In order to emphasize the role of the nonequilibrium state in the demon, we will furthermore assume all $g_l$ to be equal, such that the Hamiltonian is inversion-symmetric in the direction of transport, i.e., under the exchange $(1,A,L)\leftrightarrow(2,B,R)$. 
The qutrit-qutrit coupling ($\lambda_{02}$, $\lambda_{12}$ and $\lambda_{01}$) can also be seen as an exchange of photons~\cite{vanLoo_photon_2013,marxer_long_2023,cheung_photon_2023,dijkema_two_2023}, or even phonons~\cite{bienfait_phonon_2019}. Importantly these photons are not necessarily of the same frequency as those from the baths. 

For the numerical calculations, except where explicitly mentioned, we will fix the frequencies $\omega_s=\unit[2]{GHz}$, $\omega_d=\unit[4]{GHz}$, and (setting $\hbar=\kB=1$) temperature $T=\unit[4]{GHz}$, around \unit[30]{mK}. The currents and noise appearing in the plots are normalized by the references $\dot{Q}_0=\hbar\unit[]{GHz^2}\approx\unit[0.11]{fW}$ and $S_0=\hbar^2\unit[]{GHz^3}\approx\unit[11]{pW^2s}$, both within nowadays experimental resolution~\cite{ronzani_tunable_2018,lu_steady_2022}.

\subsection{Basic cycle: the fluctuating switch}
\label{sec:cycle}

Consider the perfectly filtered case such that each reservoir couples to a single transition and with $\lambda_{12}=\lambda_{01}=0$. The exchange of photons with L and R occur via the transitions $|1j\rangle\leftrightarrow|2j\rangle$ and $|i1\rangle\leftrightarrow|i2\rangle$, respectively, as depicted in Fig.~\ref{fig:scheme}(b). Let us assume $T_L\geq T_R$. The expected heat current from L to R is conditioned on the occupation of the states $|1\rangle_q$, which requires the demon reservoirs to have excited one qutrit at an earlier time. In this sense, the occupation of the ground states of each qutrit effectively uncouples it from the system reservoirs connected to it (L for 1, R for 2), as illustrated in Fig.~\ref{fig:cycle}.
Fluctuations by photon absorption/emission from/to one of the demon reservoirs effectively switch the coupling to one of the system ones on/off. These noisy couplings are sufficient to induce a rectification effect, which requires an asymmetry. In our case, where the Hamiltonian is symmetric, the asymmetry is introduced by the dynamics: the rates of the two switching mechanisms are different for having different temperatures, $T_A$ and $T_B$. 

Since $T_B>T_A$, the excitation of the qutrits is more likely to occur via a photon from reservoir B. This transition populates state $|1\rangle_{\ttb}$, thus effectively switching the coupling of qutrit {\ttb} and reservoir R on, and this way allowing the system to absorb a photon from reservoir R (the cold one!). The system thus reaches state $|02\rangle$. The swap transition transfers the excitation from qutrit 2 to 1, effectively switching the {\tta}-L coupling on. It also switches the {\ttb}-R coupling off, so the unlikely absorbed photon cannot go back to R. In this sense, the coherent coupling $\lambda_{20}$ acts as a turnstile that changes the reservoir to which the qutrit system is put in contact with, bearing resemblances with cyclic Otto refrigerators~\cite{segal_molecular_2006,abah_optimal_2016,karimi:2016,irene} but requiring no active driving (the cyclic behaviour in our device is purely stochastic) and with the opening of the membrane hole in Maxwell's original formulation of his demon~\cite{maxwell:book}. The analogy of Otto cycles in exchange-coupled qubits and quantum controlled Maxwell demons~\cite{lloyd_quantum_1997,kieu_second_2004,quan_maxwells_2006} has already been pointed out~\cite{quan:2007}.

State $|20\rangle$ then relaxes by emitting a photon to L (the hot one!) and subsequently to A. This resets the initial state, after a cycle as the one highlighted in Fig.~\ref{fig:cycle} by the clockwise arrows. Along the cycle, a quantum of heat $\hbar\omega_s$ has been transferred from the cold to the hot system reservoirs, at the expense of $\hbar\omega_d$ being transported from A to B. 
Other transitions are possible (represented in grey in Fig.~\ref{fig:cycle}) that involve states with both qutrits excited. However these only introduce fluctuations to the basic cycle, as transport requires the swapping of the individual qutrit ground states mediated by the coupling $\lambda_{02}$ (recall we are neglecting $\lambda_{12}$ and $\lambda_{01}$ at this point).

In the weak coupling limit and in the stationary regime, the entropy of reservoir $l$ will decrease by $\Delta\Sigma_l=\Delta Q_l/T_l$ for every amount of extracted heat $\Delta Q_l$~\cite{Strasberg2022}. The second law will favor the above sequence provided the condition
\beq
\label{eq:2ndlaw}
\omega_d\left(\frac{1}{T_A}-\frac{1}{T_B}\right)>\omega_s\left(\frac{1}{T_R}-\frac{1}{T_L}\right)
\eeq
is fulfilled. An observer with access limited to reservoirs L and R would hence measure that, while energy is conserved (the first law is respected), entropy has decreased by an amount given by minus the right-hand side of Eq.~\eqref{eq:2ndlaw}, and could interpret this as a violation of the second law (and of common sense). Of course, the larger increase of entropy in A and B [left hand side of Eq.~\eqref{eq:2ndlaw}] guarantees the global increase of entropy. 

With the condition $T_B>T_A$ fixed, Eq.~\eqref{eq:2ndlaw} imposes that the demon operation is restricted to the system temperatures being $T_L>T_R$, as a consequence of the qutrit symmetry. Different qutrit compositions result in different temperature conditions, as discussed in Sec.~\ref{sec:otherdemons}.  

Note also that in every cycle, $\Delta Q_L+\Delta Q_R=0$ and $\Delta Q_A+\Delta Q_B=0$, hence the demon condition for no heat exchange between system and demon is fulfilled by construction in the stationary regime. It does not depend on the particular configuration of system bath-couplings (as long as we stick to the weak coupling regime) nor on sets of temperatures. Note that fluctuations in the heat exchanged with the demon terminals are temporarily stored in the qutrits (in the form of an excitation that is eventually released back to a demon terminal) and do not flow into the system reservoirs (which are filtered at a different frequency) at any time.

\subsection{Partitions}
\label{sec:part}
Partitioning the system in different regions allows to interpret the dynamics under the appropriate conditions. We can for instance distinguish the measured reservoirs (L and R) and the environment of which one knows nothing (the demon baths A and B), what we call the partition AB$|$LR.
When the conditions $\dot{Q}_R=-\dot{Q}_L$ and $\dot{Q}_B=-\dot{Q}_A$ are met, which will be the case in most of the cases below (except for Sec.~\ref{sec:leakage}), we are allowed to respectively define the system and demon currents as:
\begin{align}
\dot{Q}_s=\dot{Q}_R=-\dot{Q}_L\quad\text{and}\quad
\dot{Q}_d=\dot{Q}_B=-\dot{Q}_A.
\end{align}
Note however that this distinction is only conceptual: the device is not bipartite (as is the case of state-dependent feedback demons~\cite{strasberg:2013,sanchez:2019,poulsen_quantum_2022}) in the sense that the system and demon terminals are coupled to the same system components (the two qutrits). 
The heat current $\dot{Q}_s$ induced by $\dot{Q}_d$ is also reminiscent of the thermal drag effect~\cite{bhandari:2018}, which however requires a heat transfer across the partition.

The spatial separation of the demon terminals (with each one coupled to a different qutrit) is essential for our configuration. Also, in the case where $\lambda_i\ll\hbar\omega_\alpha$, the dynamics can be described in terms of the density matrix of states of qutrits {\tta} and {\ttb} (local description). This allows for a meaningful partition AL$|$BR through which we can define the heat flow:
\begin{equation}
\dot{Q}_{c}=\dot{Q}_R+\dot{Q}_B=-\dot{Q}_L-\dot{Q}_A,
\end{equation}
which could be interpreted as the internal current, in the spirit of Ref.~\cite{freitas_2021}. 
The need to filter the system and demon transitions impose that $\omega_d\neq\omega_s$ and therefore $\dot{Q}_c$ can only vanish when all $\dot{Q}_l=0$, in this configuration. Note however that the interpretation of $\dot{Q}_c$ as an internal current is less clear for stronger couplings, where the states of the two qutrits hybridize such that the local description is no longer meaningful.

\section{Currents, noise and correlations}
\label{sec:transport}

The dynamics of the coupled qutrit system is described, in the weak qutrit-reservoir coupling limit and assuming Markov and secular approximations, by the Gorini-Kossakowski-Sudarshan-Lindblad master equation~\cite{breuer:book} of the reduced density matrix, $\dot\rho={\cal L}\rho$, with:
\begin{equation}
\label{eq:lindbladian}
{\cal L}X=-\frac{i}{\hbar}[H_S,X]+{\cal D}X,
\end{equation}
where ${\cal D}=\sum_l{\cal D}_l$ represents the dissipative dynamics induced by the reservoirs, being
\beq 
{\cal D}_lX{=}\sum_{jk}W_{jk}^{l}\big(Y_{q,jk}XY_{q,jk}^\dagger-\frac{1}{2}\{Y_{q,jk}^\dagger Y_{q,jk},X\}\big),
\eeq
and where we sum all possible jump operators $Y_{q,jk}$ describing transitions $|k\rangle_q\rightarrow|j\rangle_q$ of qutrit $q$ that are allowed by the system-bath coupling Hamiltonian introduced in Sec.~\ref{sec:model} (the qutrit index, $q$, is fixed by the involved reservoir, $l$). We assume a local master equation, valid in configurations for which $\lambda_\alpha<\Gamma_l$, see e.g. Ref.~\cite{hofer:2017njp}. For later convenience, we write the transition rates as $W_{jk}^{l}=W_{jk}^{ls}$ with $s = \mathrm{sgn} (\omega_{jk})$, to distinguish when the transition $|k\rangle_q\rightarrow|j\rangle_q$ is due to reservoir $l$ absorbing ($s=-$) or emitting ($s=+$) a photon of frequency $\omega_{jk}$, with the Fermi golden rule form:
\begin{equation} 
W_{jk}^{ls}=s\Gamma_l\zeta_{l}(|\omega_{jk}|,z_l)n_l(\omega_{jk}),
\label{eq:Wjk}
\end{equation}
with $\Gamma_l\propto|g_l|^2$
and the Bose-Einstein distribution function describing the occupation of reservoir $l$:
\begin{equation}
n_l(\omega)=\left[\exp(\hbar\omega/\kBT_l)-1\right]^{-1}.
\end{equation}
The system-bath couplings are filtered at different frequencies $\omega_l$ (with $\omega_{R,L} = \omega_s$ and $\omega_{A,B} = \omega_d$), which we assume to have a Lorentzian shape
\beq
\label{eq:filter}
\zeta_{l}(\omega,z_l)=\frac{z_l^2}{(\omega-\omega_l)^2+z_l^2}
\eeq
of width $z_l$, representative of resonator mediated couplings~\cite{pekola_colloquium_2021}. We consider symmetric filters such that $z_l = z_r \; \forall l$, and through most of the paper we will assume perfect filtering such that $z_r\rightarrow0$. 

The heat currents and their correlations are calculated in the stationary regime defined by $\dot\rho=0$. In the spirit of the full counting statistics approach~\cite{levitov_charge_1993}, we express the heat transport in terms of the distribution of the number of photons of different frequencies $\omega_\alpha$ absorbed by reservoir $l$, i.e., $N_{l\alpha}$, each of them carrying an amount of heat $\hbar\omega_\alpha$. The first two moments of the distribution (mean and variance) give the photon currents
\begin{equation}
\label{eq:IN}
I_{l\alpha}=\frac{d}{dt}\langle N_{l\alpha}\rangle
\end{equation}
and their auto- ($l=l'$) and cross-correlations ($l\neq l'$):
\begin{equation}
\label{eq:corrN}
S_{l\alpha,l'\beta}^N=\frac{d}{dt}\left(\langle N_{l\alpha}N_{l'\beta}\rangle-\langle N_{l\alpha}\rangle\langle N_{l'\beta}\rangle\right).
\end{equation}
With these we obtain the heat currents
\begin{equation}
\dQ_l=\hbar\sum_\alpha\omega_\alpha I_{l\alpha}
\label{eq:dotQ}
\end{equation}
and the heat noise correlators
\begin{equation}
S^Q_{ll'}=\hbar^2\sum_{\alpha\beta}\omega_\alpha\omega_{\beta}S^N_{l\alpha,l'\beta}.
\label{eq:SQll}
\end{equation}
Details of the derivation and full expressions for the correlations are given in Appendix~\ref{sec:fcs}. In what follows, we will drop the superscript and refer the heat current correlations as $S_{ll'}$.

\section{Performance quantifiers}
\label{sec:performance}

With the heat currents and noises we can characterize the performance of the device as follows.

\subsection{Efficiencies}
\label{sec:effs}

\subsubsection{Free energy efficiency}
\label{sec:freeeff}

Multibath systems have the possibility to perform multiple tasks simultaneously~\cite{entin:2015,manzano:2020,lopez_optimal_2022,hammam_quantum_2023} by using more than one resource~\cite{jiang_enhancing_2014,lu:2017,lu:2019,fatemeh,manzano:2020}. Our case here complies with both possibilities in the warm demon condition: the coldest bath (L) is cooled and heat is pumped into the hottest one (R) by using heat flowing in the two demon baths. In such a case, one needs to generalize the definition of efficiency so it includes multiple operations and resources~\cite{manzano:2020}.
For this, it is useful to consider the changes of the generalized free energies $\dot{F}_l=\dot{Q}_l-T_0\dot\Sigma_l$, which give the maximal amount of work than a bath at temperature $T_0$ can extract from the different reservoirs~\cite{fatemeh,manzano:2020,tesser_thermodynamic_2023}. The temperature $T_0$ can be considered as the ambient temperature from the perspective of non-equilibrium reservoirs~\cite{fatemeh} or as a reference temperature from an operational point of view~\cite{manzano:2020}. In our case, it is natural to assign it to the equilibrium temperature $T_0=T$. The stationary entropy production rate reservoir $l$, in the weak coupling limit is $\dot\Sigma_l=\dot Q_l/T_l$.  Then, noticing that $\dot{F}_l=(T_l-T)\dot\Sigma_l$, free energy is generated in the reservoirs which, being at a higher/lower temperature than $T$, nevertheless absorb/emit heat.
The efficiency is then defined in terms of the contribution of the terminals where free energy is being generated (the system) over those where it is consumed (the demon reservoirs acting as resources):
\begin{equation}
\eta_f=\frac{\dot{F}_s}{-\dot{F}_d}=\frac{\dot{F}_L+\dot{F}_R}{-\dot{F}_A-\dot{F}_B},
\label{eq:eta_f}
\end{equation}
Note however that under the demon condition, $\dot{Q}_d=\dot{Q}_s=0$, the efficiency is independent of $T_0$ and simply reads:
\begin{equation}
\eta_f=\frac{-\dot\Sigma_s}{\dot\Sigma_d}=\frac{-\dot\Sigma_L-\dot\Sigma_R}{\dot\Sigma_A+\dot\Sigma_B},
\end{equation}
in terms of the system and demon entropy production rates. It then coincides with entropic efficiencies~\cite{lu_multitask_2022}, see also Ref.~\cite{strasberg_clausius_2021}.
The interpretation is clear: an efficient demon will be one that generates as much entropy as is reduced in the system (or almost).

\subsubsection{Heat efficiencies}
\label{sec:heateff}
It will also be useful to consider more conventional efficiencies defined in terms of heat absorbed from the hot demon bath to achieve a given operation, e.g, cooling:
\begin{equation}
\label{eq:heateff}
\eta_h=\frac{\dot{Q}_R}{\dot{Q}_B},
\end{equation}
though nothing prevents one to consider bath A as a resource, instead.

The analogous efficiency taking into account that the thermodynamic resource is composed by two baths 
\beq
\eta_{AB}=\frac{\dot{Q}_R}{|\dot{Q}_A+\dot{Q}_B|},
\eeq
is not bounded and diverges under the demon condition $\dot{Q}_A+\dot{Q}_B=0$, which gives information on when the device is working beyond the thermodynamic bounds, defined by some bound. In the case $\dot{Q}_A+\dot{Q}_B\neq0$, when there is heat leaking from the demon, we set the bound by comparing with the case where the demon is replaced by a single thermal bath, call it E, at temperature $T$: $\tilde\eta_0\equiv T_R(T_L-T)/T(T_L-T_R)$, see Appendix~\ref{sec:3teff} for details. Then, the demonic effect manifests when
\beq
\label{eq:relaxdem}
\eta_{AB}>\tilde\eta_0,
\eeq
emphasizing the nonequilibrium state of the demon reservoirs.
In particular, one can define the operation of a {\it relaxed} demon that allows for a finite exchange of heat with the system~\cite{fatemeh}. In our case, this happens when the system-bath couplings are not ideal filters, see Sec.~\ref{sec:leakage}.

\subsection{Fluctuations}
\label{sec:fluct}

\subsubsection{Thermodynamic uncertainty relations}
\label{sec:tur}

To quantify the presence of noise in the cooling power, we use the thermodynamic uncertainty relation which introduces a bound for the minimal fluctuations of a thermodynamic output current in classical Markovian systems. We use it to quantify the precision of the cooling power, $\dot{Q}_R$, via the coefficient:
\beq
\label{eq:QTUR}
{\cal Q}=\frac{\dot\Sigma S_{RR}}{\dot{Q}_R^2}\geq2,
\eeq
such that very noisy currents give $2/{\cal Q} \ll 1$, while $2/{\cal Q}\rightarrow1$ saturates the classical bound. This relation has been predicted to be violated in quantum coherent devices~\cite{agarwalla_assessing_2018,saryal_thermodynamic_2019}, in particular in qutrit masers~\cite{kalaee_violating_2021,singh_thermodynamic_2022}. However, this occurs at larger couplings than the ones we are interested in in the present work.
We note also that these relations have been extended to multiple reservoir systems~\cite{dechant_multidimensional_2018,lopez_optimal_2022} which could be tested in our setup. We are not worried about these details here, and will simply use the classical version \eqref{eq:QTUR} to quantify the device performance.

\subsubsection{Pearson coefficient}
\label{sec:pearson}

We are particularly  interested in the crosscorrelation of currents in the system terminals. The Pearson coefficient 
\begin{equation}
\label{eq:pearson}
\epsilon_P=
\frac{S_{LR}}{\big(S_{LL}S_{RR}\big)^{1/2}},
\end{equation}
measures the crosscorrelation of the system currents with respect to the corresponding autocorrelations. Maximally correlated currents give $\epsilon_P=\pm1$, a proposed criterium for the {\it strict Maxwellianity} of autonomous demons including internal currents~\cite{freitas_2021}.
Note that currents of different nature can also be maximally correlated: emitted electrons and photons in a quantum dot~\cite{sanchez:2008} or charge and heat currents in Coulomb coupled conductors~\cite{sanchez:2013}. 
In our case, the two currents $\dot{Q}_L$ and $\dot{Q}_R$ may be related by an internal current (given by photons exchanged via the coherent couplings $\lambda_{02}$ and $\lambda_{12/01}$). However there is not a continuous flow between L and R: the internal current corresponds to photons of different frequency than those exchanged with the reservoirs.

\section{Demonic behaviour: Heat flows from cold to hot}
\label{sec:demon}

\begin{figure}[t]
\includegraphics[width = \linewidth]{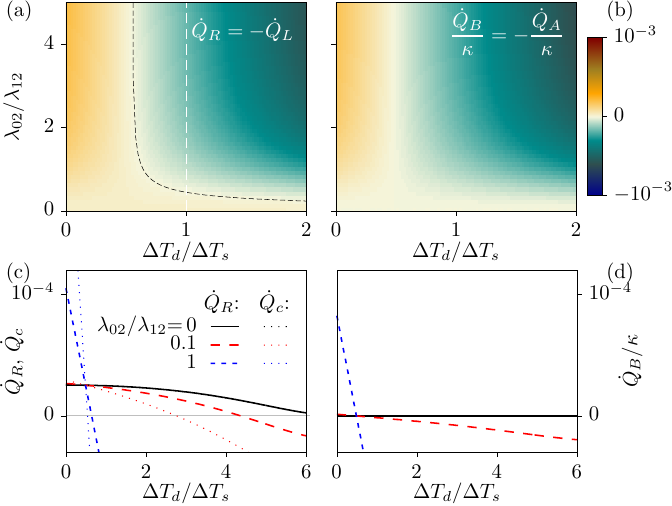}
  \caption{Conditions for the demon. Heat current into reservoirs (a) R and (b) B as a function of the temperature difference $\Delta T_d$ and the coupling $\lambda_{02}$. Parameters (in GHz): $\Gamma_l=\Gamma = 0.01$, $\omega_s=2$, $\omega_d=4$, $T=4$, $\lambda_{12}=0.01$, $\lambda_{01}=0$ and $\Delta T_s=1$. The black dashed line marks the vanishing of the current in R. In the region between the black and the white dashed lines in (a), the resource terminals are warm with respect to the system ones and cause cooling, nevertheless. The demon quantities are divided by $\kappa=\omega_d/\omega_s$, for an easier comparison between panels. (c), (d) Cuts of the previous for different values of the couplings $\lambda_{02}$. The thin dotted lines in (c) correspond to the internal current $\dot{Q}_c$.}
    \label{fig:map}
\end{figure}

Once we have the ingredients to compute the transport properties, we start by describing how the expected heat currents in the system are reversed by coupling to the demon baths. The configuration for which the effect of the demon is minimal is when $\lambda_{12}\neq0$ and $\lambda_{01}=\lambda_{02}=0$. Then, heat flows between L and R via transitions between the effective two level systems formed by states $|1\rangle_q$ and $|2\rangle_q$, as in a system of coupled qubits~\cite{kargi_quantum_2019,xu_heat_2021,khandelwal_critical_2020,iorio_photonic_2021}, with the difference that the transport states need to be populated by photons from reservoirs A and B (noise-induced transport regime). Assuming homogeneous couplings $\Gamma_l=\Gamma$, we get 
\begin{equation}
\dot{Q}_R=\hbar\omega_s\lambda_{12}^2\Gamma p(\{T_l\},\{\omega_l\})[n_L(\omega_s)-n_R(\omega_s)],
\end{equation}
shown in Fig.~\ref{fig:map}(a) in the region $\lambda_{02}\ll\lambda_{12}$.
The demon reservoirs hence act as switches: they affect the relative occupation of the transport states via the prefactor $p(\{T_l\},\{\omega_l\})\propto n_A(\omega_d)n_B(\omega_d)$, but do not inject any heat, see Fig.~\ref{fig:map}(b). In that case, for a given temperature difference $\Delta T_s>0$ heat flows from L (hot) to R (cold), $\dot{Q}_R<0$, as expected, irrespective of the situation at terminals A and B, see Fig.~\ref{fig:map}(c). As $\Delta T_d$ increases, such that the temperature of reservoir A decreases, current is suppressed as $e^{-\hbar\omega_d/\kBT_A}$, in a sort of dynamical channel blockade~\cite{cottet_positive_2004}. For $\kBT_A\ll\hbar\omega_d$, thermal fluctuations can rarely excite qutrit {\tta} and transport is switched off.

As the coupling $\lambda_{02}$ becomes finite, heat flows at reservoirs A and B, see Figs.~\ref{fig:map}(b) and \ref{fig:map}(d). For a high enough temperature difference $\Delta T_d$, the heat flow in the system reservoirs is reversed: qutrit {\tta} works as an absorption refrigerator and qutrit {\ttb} as a heat pump~\cite{geusic:1967}. By construction, both currents are equal in magnitude $\dot{Q}_L+\dot{Q}_R=0$, hence the entropy production rate \begin{equation}
\label{eq:entrops}
\dot\Sigma_s\equiv\dot\Sigma_L+\dot\Sigma_R=\dot{Q}_R\eta_C/T_R
\end{equation}
is negative in the system. Here, $\eta_C=1-T_R/T_L$ is the Carnot efficiency. Hence, while the first law of thermodynamics is fulfilled both in the system and demon regions, the second law is violated in the system region when $\dot{Q}_R>0$. Notably, this occurs even if $\Delta T_d<\Delta T_s$, such that the coldest reservoir is cooled further and heat is pumped into the hottest one (the {\it warm demon}), see the region limited between dashed lines in Fig.~\ref{fig:map}(a). 

However, the reversal of heat currents in the system reservoirs is not necessarily accompanied by a reversal of the internal flow, plotted with dotted lines in Fig.~\ref{fig:map}(c): $\dot{Q}_c$ and $\dot{Q}_R$ are both positive only in a limited region. The size of this region increases with $\lambda_{02}/\lambda_{12}$ and will occupy all the phase space for $\lambda_{02}/\lambda_{12}\rightarrow\infty$ which is the case that we discuss in detail in Sec.~\ref{sec:perfectfilters}. 
In the opposite regime, $\Delta T_d\ll\Delta T_s$, it is the demon currents which are reversed, see Fig.~\ref{fig:map}(b). Differently from the system currents, the temperature difference at which $\dot{Q}_B$ changes sign is independent of $\lambda_{02}$, see Fig.~\ref{fig:map}(d). Of course, the total entropy never decreases: $\dot\Sigma=\sum_l\dot{Q}_l/T_l\ge0$.

\begin{figure}[t]
\includegraphics[width = .975\linewidth]{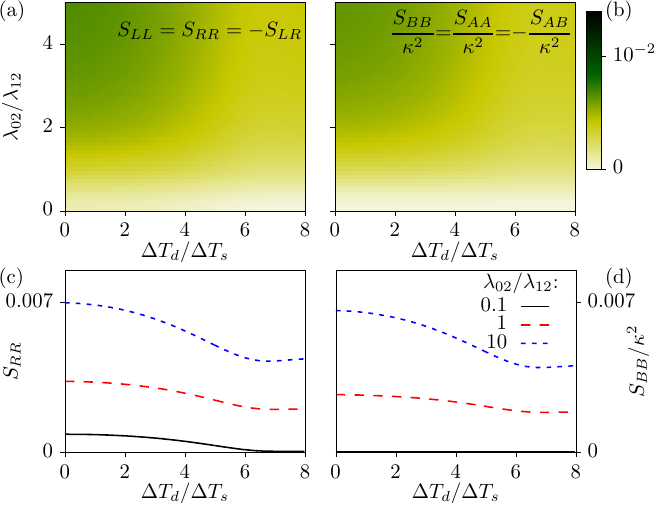}
  \caption{Heat noise in reservoirs (a) R and (b) B as a function of the temperature difference $\Delta T_d$ and the coupling $\lambda_{02}$, for the same parameters as in Fig.~\ref{fig:map}. The demon quantities are divided by $\kappa^2$, for an easier comparison between panels. (c), (d) Cuts of the previous for different values of $\lambda_{02}$.}
    \label{fig:mapnoise}
\end{figure}

The current-current correlations give additional insight, see Fig.~\ref{fig:mapnoise}. For small $\lambda_{02}$, the system noise is monotonically suppressed with $\Delta T_d$, following the behaviour of $\dot{Q}_R$. The system noise increases with the coupling $\lambda_{02}$, as it is affected by the fluctuations of the demon reservoirs. As expected, the demon reservoirs noise increases with the onset of $\dot{Q}_d$ as the coupling $\lambda_{02}$ increases, see Fig.~\ref{fig:mapnoise}(d). It eventually dominates the fluctuations of the system, compare Figs.~\ref{fig:mapnoise}(c) and \ref{fig:mapnoise}(d). However neither the system nor the demon noises have any feature related to the reversal of the heat currents. Remarkably, in each subsystem the current-current correlations are identical:
\begin{equation}
\label{eq:symnoise}
    S_{LL}=S_{RR}=-S_{LR},
\end{equation}
and similarly for reservoirs A and B, see Figs.~\ref{fig:mapnoise}(a) and \ref{fig:mapnoise}(b). Equation \eqref{eq:symnoise} is a consequence of the conservation of energy in the AB$|$LR partitions at every cyclic sequence, not only on average: a photon emitted by one bath is either absorbed by the other bath in the same partition (leading to heat transport) or reabsorbed (no transport). It also results in a Pearson coefficient $\epsilon_P=-1$, one of the requirements of a strict Maxwell demon, according to Ref.~\cite{freitas_2021}.  However, like in an N-demon~\cite{ndemon} (where conservation of energy occurs only on average), there is no clear notion of information processing nor any feedback mechanism.

\section{Ideal operation: perfectly filtered system}
\label{sec:perfectfilters}

In the following, we discuss the operation of the device for different configurations, in terms of the cooling power and efficiency, as well as the noise properties. 

Let us first discuss the optimal case in which the two qutrits are only coupled via the term proportional to $\lambda_{02}$ in Eq.~\eqref{eq:hab} i.e., $\lambda_{12}=\lambda_{01}=0$. The couplings to the baths are also perfectly filtered i.e., $z_r\rightarrow0$. This configuration corresponds to the limit $\lambda_{02}\gg\lambda_{12}$ considered in Sec.~\ref{sec:demon}. Then, the only cycle that contributes to transport is the one highlighted in Fig.~\ref{fig:cycle} and described in Sec.~\ref{sec:cycle}. The transition $|02\rangle\leftrightarrow|20\rangle$ acts as a bottleneck for the heat currents: in order to have an excitation transfer between qutrits, one of them has to have consecutively been excited by a demon and then by a system bath, in a sort of sequential (incoherent) upconversion process. The other qutrit then downconverts the excitation in photons absorbed by its two baths. In the completion of every basic cycle, one photon of frequency $\omega_s$ has been emitted by reservoir R and one absorbed by reservoir L, and correspondingly with photons of frequency $\omega_d$ by reservoirs B and A. In the stationary regime, hence 
\beq
\label{eq:identicalQl}
\frac{\dot{Q}_A}{\omega_d}=\frac{\dot{Q}_L}{\omega_s}=-\frac{\dot{Q}_B}{\omega_d}=-\frac{\dot{Q}_R}{\omega_s}.
\eeq
Therefore the system and demon currents and, as a consequence, also $\dot{Q}_c$, are all reversed at the same point.  
Assuming that the temperature increases are applied symmetrically in the system and demon baths, we find the vanishing current condition to be uniquely depending on the temperature of the demon baths:
\begin{equation}
\label{eq:Ss0}
\Delta T_s^0=\sqrt{\xi_d^2+4T^2}+\xi_d,
\end{equation}
with $\xi_d\equiv(\Delta T_d^2-4T^2)\omega_s/2\Delta T_d\omega_d$.

\begin{figure}[t]
\includegraphics[width = \linewidth]{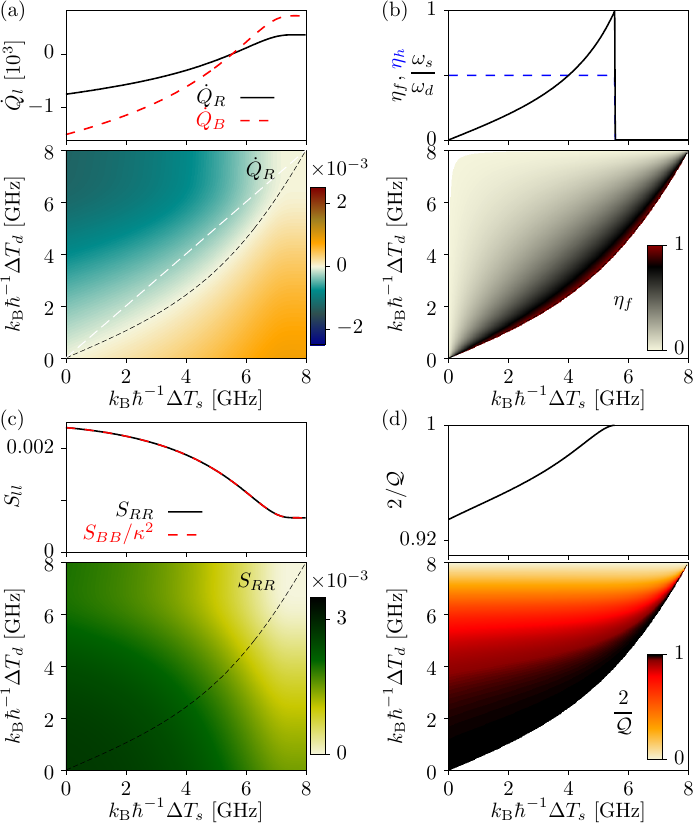} 
  \caption{Perfectly filtered case. The density plots show (a) the cooling power, (b) efficiency, (c) noise and (d) ${\cal Q}$ for heat extracted from reservoir R as functions of the temperature differences $\Delta T_s$ and $\Delta T_d$. Parameters (in GHz): $\Gamma_l=\Gamma = 0.01$, $\omega_s=2$, $\omega_d=4$, $T=4$, $\lambda_{02}=0.01$, $\lambda_{01}=\lambda_{12}=0$. White regions in (b) and (d) mark the configurations with no cooling. The dashed black line in all plots is a guide to the eye that marks the condition $\dot\Sigma_s=0$, according to Eq.~\eqref{eq:Ss0}. In the region between the black and the (diagonal) white dashed lines in (a), the resource terminals are warm with respect to the system ones. The line plots show cuts of (a) to (d), for $\Delta T_d=4$, including as well the current and noise of reservoir B in (a) and (c), for comparison, with $\kappa=\omega_d/\omega_s$. The blue-dashed line in (b) corresponds to the heat efficiency $\eta_h$.}
    \label{fig:PP02_OPTJS}
\end{figure}

Equation~\eqref{eq:Ss0} hence sets the boundary of the region where heat is extracted from reservoir R. This is plotted in Fig.~\ref{fig:PP02_OPTJS} as a function of the system and demon temperature differences. For $\Delta T_s<\Delta T_s^0$, the system works as a refrigerator induced by the heat transferred between the demon reservoirs. Remarkably, this occurs even if the demon reservoirs are closer to equilibrium (warmer) than the system ones i.e., having $\Delta T_s>\Delta T_d$. This is the region between the black-dotted and the white-dashed lines in Fig.~\ref{fig:PP02_OPTJS}(a). The {\it warm demon} operation is here possible because $\omega_d>\omega_s$. It turns out that the efficiency is also largest in this region, as shown in Fig.~\ref{fig:PP02_OPTJS}(b), with $\eta_f$ approaching 1 for $\Delta T_s\rightarrow\Delta T_s^0$. 
For $\Delta T_s>\Delta T_s^0$, the role of the system and the demon are reversed. Note that $\Delta T_s^0\rightarrow0$ for $\Delta T_d\rightarrow0$, i.e., there is no possible demon working at equilibrium. 

The same arguments that led to Eq.~\eqref{eq:identicalQl} apply to the fluctuations, leading to
\beq
\label{eq:identicalSll}
\frac{S_{AA}}{\omega_d^2}=\frac{S_{BB}}{\omega_d^2}=-\frac{S_{AB}}{\omega_d^2}=\frac{S_{LL}}{\omega_s^2}=\frac{S_{RR}}{\omega_s^2}=-\frac{S_{LR}}{\omega_s^2},
\eeq
i.e., all currents are perfectly correlated and result in a maximal Pearson coefficient $\epsilon_P=-1$. The noise of the extracted current is plotted in Fig.~\ref{fig:PP02_OPTJS}(c), showing (again) no particular feature related to the reversal of the heat currents. It monotonically decreases as one of the reservoirs gets colder. The TUR however becomes saturated as $\Delta T_s\rightarrow\Delta T_s^0$, see Fig.~\ref{fig:PP02_OPTJS}(d). At this point, not only the refrigerator has the maximal efficiency $\eta_f\rightarrow1$, but is also maximally precise, ${\cal Q}\rightarrow2$. 

We then conclude that the device performs under the conditions of a strict Maxwell demon for an experiment accessing not only the currents and fluctuations of the system terminal, L and R, but also having information on the reversal of the AL$|$BR interpartition flow.

\section{Less ideal cases}
\label{sec:lessideal}

Now we relax the ideal conditions of the previous section by considering that all couplings are mediated by finite-width filters, such that either the secondary swap transitions $|01\rangle\leftrightarrow|10\rangle$ and $|12\rangle\leftrightarrow|21\rangle$ become possible, or the qutrit-bath couplings become leaking: e.g., with the system reservoirs (L and R) being able to excite the ground state.  

\subsection{Additional exchange couplings}
\label{sec:secondary}

\begin{figure}[t]
    \centering
    \includegraphics[width = \linewidth]{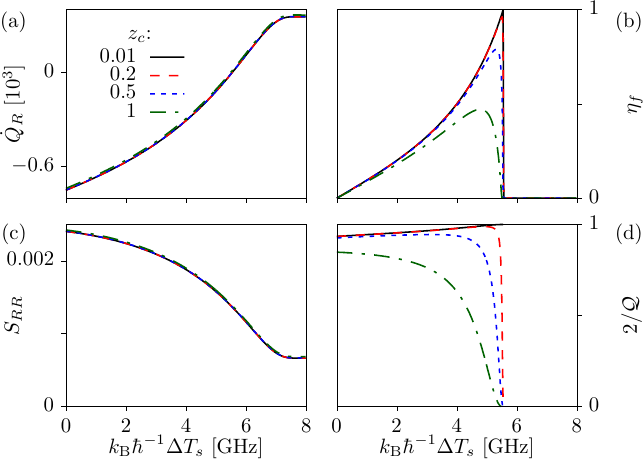}
    \caption{ Additional couplings: (a) Heat currents, (b) efficiencies, (c) noise and (d) $Q$ for heat extracted from reservoir R as functions of the temperature difference $\Delta T_s$, for different $z_{c}$ and $\Delta T_d=4$ and the same other parameters as in Fig.~\ref{fig:PP02_OPTJS}. Efficiencies and ${\cal Q}$ are plotted in the cooling region where $\dot{Q}_R>0$.}
    \label{fig:PP02SECcuts}
\end{figure}

We first explore the effect of additional swap transitions. We do this by assuming that the qutrit coupling is mediated by a resonance of width $z_{c}$  around $\omega_0 = \omega_s + \omega_d$, for which we consider
$\lambda_{\alpha}\rightarrow \lambda_{02}\zeta_{0}(\omega_\alpha,z_{c})$, using Eq.~\eqref{eq:filter}.
As the width of the resonance increases, additional couplings start to contribute. The effect on transport, plotted in Fig.~\ref{fig:PP02SECcuts}, shows that, though the cooling power is barely affected by $z_{c}$, the efficiency is reduced. The reason is the onset of cycles $|00\rangle\rightarrow|01\rangle\leftrightarrow|10\rangle\rightarrow|00\rangle$ for finite $\lambda_{01}$ which transfer heat directly from B to A without involving the system reservoirs, see Fig.~\ref{fig:PP02SECcuts}(a) and \ref{fig:PP02SECcuts}(b). 
The same argument applies to the system autocorrelations and the TUR, see Figs.~\ref{fig:PP02SECcuts}(c) and (d). 

The conservation of energy in each cycle imposes that the fluctuations in L and R, and in A and B are the same in the steady state, no matter in how many different ways the excitations are transferred between the qutrits. Hence the crosscorrelations are not affected by these additional transitions (as also discussed in Sec.~\ref{sec:demon} for a similar case), so the Pearson coefficient is still $\epsilon_P=-1$. Hence, the operation is still indistinguishable from that of a  Maxwell demon from measurements in L and R only. The analogy in this case can be done with a demon that controls only one of several holes in the membrane that separates the two system partitions~\cite{maxwell:book}.

\subsection{Leaking filters: relaxed demon}
\label{sec:leakage}

The effect of imperfectly filtered couplings between the qutrits and the reservoirs is more drastic, as it introduces the possibility that the two qutrit transitions are induced by the two reservoirs each one is coupled to. Experimentally, the couplings to the reservoirs can controllably be made lossy~\cite{maurya_ondemand_2023}. We do this here by considering finite width resonances (the same for all reservoirs, $z_r$) via $\zeta_{l}(\omega,z_r)$ in Eq.~\eqref{eq:filter}. 

\begin{figure}[t]
    \includegraphics[width = \linewidth]{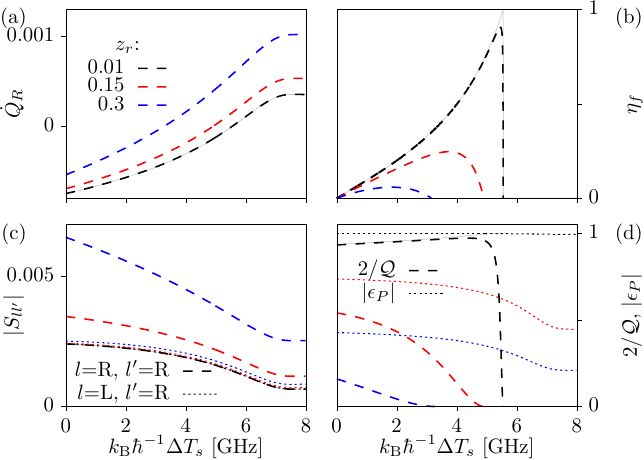}
    \caption{Effect of imperfect filtering: (a) Heat currents, (b) efficiencies, (c) auto- (dashed) and cross-correlations (dotted) of the system currents, and (d) thermodynamic uncertainty ${\cal Q}$ (dashed) and $\epsilon_P$ (dotted) for heat extracted from reservoir R as functions of the temperature difference $\Delta T_s$, for different $z_{r}$ and $\Delta T_d=4$ and the same other parameters as in Fig.~\ref{fig:PP02_OPTJS}. Efficiencies and ${\cal Q}$ are plotted in the cooling region where $\dot{Q}_R>0$. Grey lines mark the perfect filtering case as references.}
    \label{fig:PP02LEAKdemon}
\end{figure}

The additional transitions are detrimental for cooling power, as shown in Fig.~\ref{fig:PP02LEAKdemon}(a), therefore the efficiency is strongly reduced even for small $z_{r}$, see Fig.~\ref{fig:PP02LEAKdemon}(b). Also the auto- and cross-correlations are affected, breaking the symmetry of Eq.~\eqref{eq:symnoise}, cf. Fig.~\ref{fig:PP02LEAKdemon}(c), making the TUR deviate from ${\cal Q}=2$, cf. Fig.~\ref{fig:PP02LEAKdemon}(d). Most important for our discussion, the lack of perfect correlations reduce the Pearson coefficient as $z_r$ increases, see dotted lines in Fig.~\ref{fig:PP02LEAKdemon}(d).

\begin{figure}[t]
    \centering
\includegraphics[width = \linewidth]{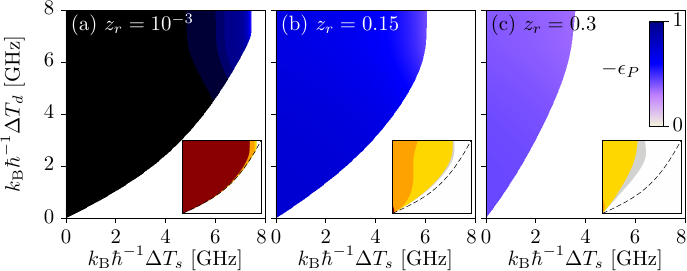}
 \caption{Pearson coefficient as a function of the temperature differences $\Delta T_s$ and $\Delta T_d$ for the case of leaking bath filters with (a) $z_r=10^{-2}$, (b) $z_r=0.15$ and (c) $z_r=0.3$. Other parameters as in Fig.~\ref{fig:PP02_OPTJS}. We only plot the relaxed demon region as defined by the condition \eqref{eq:relaxdem}. The insets are maps of the operation regions according to the same condition: grey is where the device is a nondemonic refrigerator, gold marks the relaxed demon territories with $\eta_{AB}>\tilde\eta_0$, while orange and dark-red are for increasingly efficient demons $\eta_{AB}>10$ and 100, respectively. Note that $\tilde\eta_0=(1-\Delta T_s/2T)/2\leq1/2$. The dashed black line limits the refrigeration region in the perfectly filtered case (strict demon). }
    \label{fig:PP02LEAKpears}
\end{figure}

Furthermore in this case, the demon condition $\dot{Q}_d=0$ is not necessarily fulfilled, so the first law is not verified in the system. However we can still find a {\it relaxed} demon behaviour as long as the cooling power is larger than $\dot{Q}_d$ (representing the deviation from the first law in the system induced by heat leaking from the demon), as defined by Eq.~\eqref{eq:relaxdem}, see Ref.~\cite{fatemeh} for a related behaviour. We plot the Pearson coefficient in the relaxed demon region in Fig.~\ref{fig:PP02LEAKpears}. Increasing widths $z_{r}$ not only reduce $|\epsilon_P|$, they also reduce the region where the relaxed demon operates. The map of the region territories is also shown in the insets of Fig.~\ref{fig:PP02LEAKpears}, with different colors marking the degree of violation of the efficiency ``bound" $\eta_0$. In the grey regions, the device behaves as a conventional (in the sense of nondemonic) multibath refrigerator.

\section{Alternative configurations}
\label{sec:otherdemons}

\begin{figure}[t]
\includegraphics[width=\linewidth]{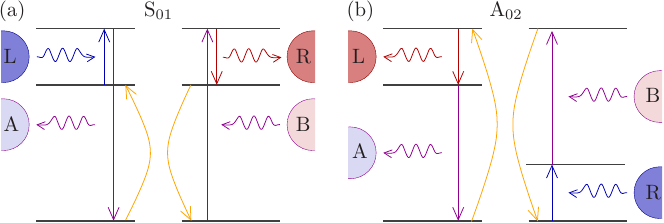}
\caption{\label{fig:schemeothers}
  Schemes of alternative configurations. (a) S$_{01}$ consists on the same qutrit disposition as for S$_{02}$ but with the demon reservoirs being filtered at a different frequency, $\hbar\omega_d=E_{2q}-E_{0q}$. The dominant exchange coupling is in this case $\lambda_{01}$. (b) A$_{02}$ considers qutrit {\ttb} to have a negative and opposite anharmonicity, with the same filters as the S$_{02}$ configuration. Note that the S$_{01}$ demon works for the opposite temperature distribution in the system, cooling L and pumping R.
}
\end{figure}

The coupling between two qutrits may have different properties depending on the experimental realization. For this reason, we explore other configurations where similar effects can be found. In the previous sections we have discussed a symmetric configuration based on two  identical qutrits with a dominant swap of the $|02\rangle$ and $|20\rangle$ states, which we name S$_{02}$ in the following. Additionally we identify two other cases which make the non-continuity of the internal current more explicit, as we will see: 
The S$_{01}$, where the two qutrits are also identical, but they are copuled via swapping the $|01\rangle$ and $|10\rangle$ states, see Fig.~\ref{fig:schemeothers}(a); and the A$_{02}$, where the system is antisymmetric for the qutrits have opposite anharmonicity, with the main coupling mediated again by $\lambda_{02}$, see Fig.~\ref{fig:schemeothers}(b). 

\subsection{Case S$_{01}$: Antiparallel internal and system currents}

The configuration S$_{01}$ needs that the demon reservoir couplings are filtered at a frequency $\omega_d^{\text{S}_{01}}=(E_{2q}-E_{0q})/\hbar$ (for the numerical calculations of this case we fix $\omega_s$=\unit[2]{GHz}). This way, baths L/R induce transitions between $|1\rangle_{{\tta}/{\ttb}}$ and $|2\rangle_{{\tta}/{\ttb}}$, and A/B between $|0\rangle_{{\tta}/{\ttb}}$ and $|2\rangle_{{\tta}/{\ttb}}$, see Fig.~\ref{fig:schemeothers}(a). Additionally the swap transition is centered around $\omega_d^{\text{S}_{01}} - \omega_s$, so that $\lambda_{01}\gg\lambda_{02},\lambda_{12}$.

The mechanism of the S$_{01}$ and S$_{02}$ systems is similar: they are both based on the asymmetric fluctuations of the {\tta}-L and {\ttb}-R switches, as discussed in Sec.~\ref{sec:cycle}, with the difference that now the demon reservoirs induce transitions between the $|0\rangle_q$ and $|2\rangle_q$ states. Hence, the cycle $|00\rangle\rightarrow|02\rangle\rightarrow|01\rangle\leftrightarrow|10\rangle\rightarrow|20\rangle\rightarrow|00\rangle$, sketched in Fig.~\ref{fig:schemeothers}(a), leads to $\hbar\omega_s$ quanta being absorbed from L and emitted into R, i.e., the role of L and R are exchanged in the cooling/pumping operation as compared to configuration S$_{02}$: they occur when $T_R>T_L$. 
This goes with an energy $\hbar(\omega_d^{\text{S}_{01}}-\omega_s)$ flowing in the opposite direction through the AL$|$BR partition via the swap transition, i.e. the system currents are antiparallel to $\dot{Q}_c$. Furthermore, in this case the presence of an internal system  current is compromised by the fact that photons are absorbed by R {\it before} they are emitted from L. 

The S$_{01}$ case can be seen as a single qutrit absorption refrigerator coherently coupled to a heat source that contains the BR partition: B and R reservoirs are both hot and inject heat into the colder partition AL (remember $T_R>T_L$ and $T_B>T_A$), which results in L being cooled down, as happens in three bath configurations~\cite{geva_three_1994}. However, the demonic effect manifest in that, at the same time, heat is being pumped into the hottest reservoir R, which is not possible with three reservoirs only.

\subsection{Case A$_{02}$: The rattrap demon}
\label{sec:rattrap}

\begin{figure}[t]
\includegraphics[width=0.7\linewidth]{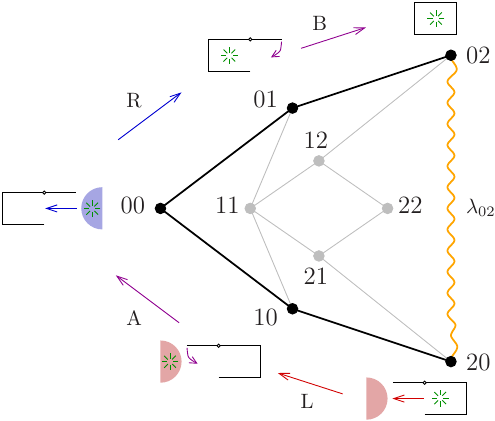}
\caption{\label{fig:rattrap}
  The rattrap demon (model A$_{02}$). Solid lines indicate the transitions  between the different states of the system. As in Fig.~\ref{fig:cycle}, black lines emphasize the basic-cycle transitions with a single excited qutrit, though the transitions involve different reservoirs, as indicated over the colored arrows. Clockwise circulation carries a photon from the cold bath R to the hot L. The orange wavy line indicates the primary coherent swap transition between states $|02\rangle$ and $|20\rangle$ with coupling $\lambda_{02}$.  
}
\end{figure}

The mechanism for the A$_{02}$ cycle, sketched in Figs.~\ref{fig:schemeothers}(b)  and \ref{fig:rattrap}, has a different interpretation. When the device is in the ground state, it is in contact with reservoirs R and A, both cold. The smaller frequency $\omega_s$ favors that a photon is absorbed from R as long as also $\omega_s/T_R<\omega_d/T_A$. Once in state $|01\rangle$, absorbing a photon from B effectively uncouples reservoir R. In this sense, the transition $|1\rangle_{\ttb}\rightarrow|2\rangle_{\ttb}$ acts as closing a trapdoor (in analogy with the trapdoor model introduced by Smoluchowski~\cite{Smoluchowski1912}) that avoids the excitation extracted from R to be released back. The excitation is hence trapped (or secured) until the swapping $|02\rangle\rightarrow|20\rangle$ transfers it to the AL partition. The {\tta}-L coupling is hence set on and the emission of a photon of energy $\omega_s$ into the hot reservoir L is enabled. The subsequent relaxation back to the ground state acts as a second trapdoor closing the 1-L coupling and resets the cycle again. 

Note that in this case, one can introduce a notion of information as done in the quantum dot models~\cite{strasberg:2013,sanchez:2019}: the excitation absorbed from reservoir R is detected by reservoir B, which emits a photon and furthermore introduces backaction by uncoupling reservoir R. The other demon terminal (A) is then used to erase the information. 

\subsection{Comparison of the models}
\label{sec:compare}

\begin{figure}[t]
    \centering
    \includegraphics[width = \linewidth]{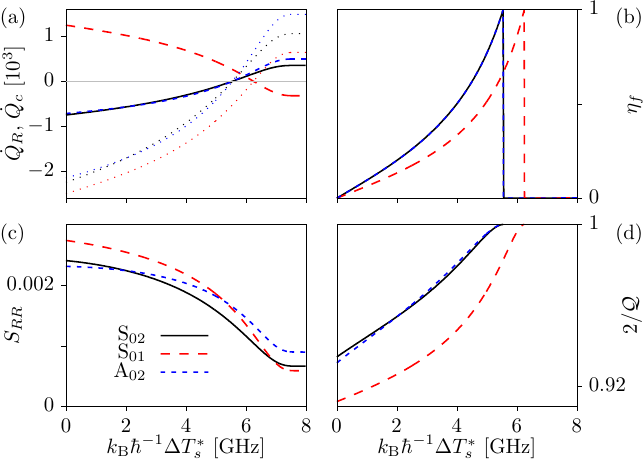}
    \caption{ Comparison of the three different configurations. (a) Heat currents, (b) efficiencies, (c) noise and (d) $Q$ for heat extracted from reservoir R as functions of the temperature difference $\Delta T_s^*=\Delta T_s$ (for the S$_{02}$ and A$_{02}$ cases) and $\Delta T_s^*=-\Delta T_s$ (for the S$_{01}$ case). The device parameters are as in Fig.~\ref{fig:PP02_OPTJS}, except for the case S$_{01}$, where $\omega_d=6$, $\lambda_{12}=0.01$ and $\lambda_{02}=\lambda_{01}=0$. The curve for S$_{02}$ is the same plotted there. The small dotted lines in (a) show the internal current, $\dot{Q}_c$ in each case.}
    \label{fig:compare}
\end{figure}

We compare the performance of the three cases in Fig.~\ref{fig:compare}, always in the ideal case with perfect filtering and only one exchange interaction (the dominant one in each case).
Note that while the demon and system currents flow in the same direction for the S$_{02}$ and A$_{02}$ cases, the opposite happens in the case S$_{01}$, see Fig.~\ref{fig:compare}(a), as expected. However, in all three cases, $\dot{Q}_c$, which is dominated by the demon flows, is reversed at the onset of the demon operation, as a consequence of the perfect correlation of the transitions. In spite of this, in the S$_{01}$ setup the cooling power and the internal current flow in opposite directions, emphasizing the lack of a continuous current in the device.

While S$_{02}$ and A$_{02}$ configurations seem to have very similar performance in terms of output power, the case S$_{01}$ has the advantage that the warm demon region is larger, because of the larger ratio $\omega_d/\omega_s$ in that configuration. This also extends the temperature region where the demon operates, as evidenced by Figs.~\ref{fig:compare}(b). The noise features are similar in all three cases, though the S$_{01}$ case is noisier in the operating region, see Figs.~\ref{fig:compare}(c) and \ref{fig:compare}(d).

\section{Discussion}
\label{sec:conclusions}

To conclude, we have proposed the coherent coupling of two qutrits in a four-reservoir configuration to reverse the heat current in the (sub)system formed by two of the baths (hence locally violating the second law of thermodynamics) without injecting any heat from the other reservoirs (hence locally respecting the first law) in experimentally relevant configurations~\cite{aamir_thermally_2023}. Thus the device operates as an autonomous demon for an observer measuring only currents and fluctuations in the system terminals. For this it requires that the two demon terminals are out of equilibrium and that the qutrit-reservoir couplings are properly filtered. The simultaneous cooling and heat pump operations in the system reservoirs does not depend on the particular configuration of the device and occurs even if the temperature difference in the demon (the driving thermodynamic force) is smaller than that of the rectified system. 

Our device has similarities and differences with other proposed models for autonomous demons based on few-level configurations. Different from most proposals~\cite{strasberg:2013,whitney:2016,ptaszynski:2018,sanchez:2019,poulsen_quantum_2022}, the device is not bipartite i.e. the state of the qutrit system cannot be separated into system and demon states, they are rather intertwined in the state space. Different from N-demons based on the coupling to nonequilibrium environments~\cite{ndemon,fatemeh}, the nonequilibrium state introduced by the demon reservoirs is spatially separated, with each bath being coupled to a different qutrit. This separation is essential to our proposal, as it takes the (necessary to a rectifier) asymmetry to the fluctuations, while all system-bath couplings can in principle be equal. 
Such a geometry was argued in Ref.~\cite{freitas_2021} to possibly induce apparent reversals of the current in two terminals of a classical multiterminal conductor by {\it non-demonic} means, which would be revealed by the properties of the current-current correlations and by the presence of an internal current that is not reversed together with the one measured in the reservoirs. In particular, a strict (autonomous) Maxwell demon is expected to show perfectly cross-correlated currents in the system baths, with a Pearson coefficient $|\epsilon_P|=1$.

We have explored the performance of the device as a simultaneous refrigerator and heat pump, focusing on both the mean heat currents and the auto and crosscorrelations in the system reservoirs, where the measurements occur. We find in the first term that 
an internal heat current can be identified by conservation of heat in the connection between qutrits (across the AL$|$BR partition). In ideal configurations (with perfectly filtered coupling to the leads and a single contribution to the qutrit-qutrit interaction) it is reversed with the system current. However this may happen at different conditions under the contribution of additional swap couplings. On top of it, the internal current is not necessarily parallel to the cooling power, as demonstrated in related configurations (S$_{01}$). Secondly, we note that a continuous current cannot be defined to flow between the system terminals and through the device: the transfer between qutrits is due to a sequence of upconversion and downconversion of the photons absorbed and emitted from/to the reservoirs. Despite these, the currents in the two system terminals are perfectly correlated via the internal dynamics in perfectly filtered configurations, even in the presence of several qutrit-qutrit couplings. Furthermore, this happens for a wide region of temperatures, not depending on the particular configuration of the system-bath couplings.

As a consequence, the mechanism cannot be distinguished from that of a {\it strict} Maxwell demon by an observer with limited information of the system. Furthermore, our device questions the relevance of an internal current: differently from conductors with classical dynamics, where all currents can be defined locally, the coherent coupling between qutrits makes the response nonlocal, avoiding the detection of the system internal dynamics (hence any internal flow) without disturbing it.

For the computation of the currents and correlations, we have extended the method of Ref.~\cite{franz} (initially introduced for charge currents in quantum dot systems) for the computation of the counting statistics of multimode and multiterminal bosonic few-level systems described by weak-coupling master equations. We use it to calculate the heat currents and the auto and crosscorrelations, and to estimate the performance of the device and its precision via the free-energy efficiency and the thermodynamic uncertainty relation. We find that the hybrid refrigerator-heat pump operation reaches the maximal efficiency at the current reversal condition (i.e., when no power is extracted) with an optimal precision saturating the classical TUR. The performance is however affected by having additional swap couplings, which reduces the efficiency and the TUR but does not affect the Pearson coefficient. Most critical is the effect of imperfect filtering in the qutrit-reservoir couplings. Not only it perturbs all performance quantifiers ($\eta_f$, ${\cal Q}$) and $\epsilon_P$: it also introduces heat leakage that compromises the conservation of heat in the system terminals: the demon conditions are in that case only met for particular temperature configurations. However, it still works as a relaxed demon when the cooling power is larger than the leaking heat across the AB$|$LR partition. 

Note that related configurations have been recently proposed for entanglement generation~\cite{tavakoli_heralded_2018} as well as for discussing synthetic negative temperatures~\cite{bera_steady_2023}, due to population inversion (an effect intimately related to the mechanism discussed here), emphasizing with our work and state of the art experiments~\cite{aamir_thermally_2023} the relevance of multibath coupled qutrits for the control of heat in quantum devices and nonequilibrium processes in quantum thermodynamic settings.

\acknowledgements
We thank S. Gasparinetti for useful discussions and M. Acciai, J. Monsel and J. Splettstoesser for relevant comments on the manuscript. We acknowledge funding from the GEFES Research Awards for Undergraduate Students 2022, the Ram\'on y Cajal program RYC-2016-20778, and the Spanish Ministerio de Ciencia e Innovaci\'on via grants No. PID2019-110125GB-I00 and PID2022-142911NB-I00, and through the ``Mar\'{i}a de Maeztu'' Programme for Units of Excellence in R{\&}D CEX2023-001316-M.

\appendix

\section{Photonic full counting statistics}
\label{sec:fcs}

To compute the heat currents and noises, we adopt a full counting statistics approach following the method of Ref.~\cite{franz}, originally developed for counting electrons emitted into one terminal of a mesoscopic conductor, and extending it to frequency-resolved particle counting in multiple reservoirs, see also Refs.~\cite{sanchez_electron_2008,spincorr,hussein:2012}. This allows us to recursively calculate the cumulants of the statistics of the number of photons  with frequency $\omega_\alpha$ that are absorbed by reservoir $l$, $N_{l\alpha}$. In an undriven system, we can relate it with the amount of heat transferred in a given time: $\langle\Delta Q_l\rangle=\sum_\alpha\hbar\omega_\alpha\langle\Delta N_{l\alpha}\rangle$. This way, we compute the statistics of both particle and heat currents from those of $N_{l\alpha}$. Here we will focus on the mean currents: 
\begin{equation}
\label{eq:IN}
I_{l\alpha}=\frac{d}{dt}\langle N_{l\alpha}\rangle
\end{equation}
and their correlations:
\begin{equation}
\label{eq:corrN}
S_{l\alpha,l'\beta}^N=\frac{d}{dt}\left(\langle N_{l\alpha}N_{l'\beta}\rangle-\langle N_{l\alpha}\rangle\langle N_{l'\beta}\rangle\right).
\end{equation}
The expressions for heat currents are obtained by replacing each $N_{l\alpha}$ by $\hbar\omega_\alpha N_{l\alpha}$ in the above expressions \eqref{eq:IN} and \eqref{eq:corrN}.

We extend the total density operator, $\rho_{S{+}B}^{}$, by introducing the vector $\chi$ whose components are the counting fields $\chi_{l\alpha}$. The resulting operator is then reduced to the system degrees of freedom by tracing out the reservoirs: ${\cal F}(\chi,t)=\Tr_B(\e^{i\chi N}\rho_{{S}{+}{B}}^{})$. Note that here $N$ is a vector containing the different particle number operators, $N_{l\alpha}^{}$.  
In the weak qutrit-bath couplings regime and assuming a secular Born-Markov approximation~\cite{Schaller2014,Strasberg2022}, we get a generalized master equation:
\begin{equation}
\label{eq:dotF}
\dot{\cal F}(\chi,t)=[{\cal L}+\sum_{l\alpha s}\left(\e^{si\chi_{l\alpha}^{\ }}-1\right){\cal J}_{l\alpha}^s]{\cal F}(\chi,t),
\end{equation}
with the index $s=\pm$ accounting for processes where a photon is absorbed by or emitted from a reservoir. It includes the usual Lindblad superoperator ${\cal L}$, including the system Hamiltonian $H_S$  and the transition rates $W_{jk}^{l\alpha s}$ given by Eqs.~\eqref{eq:lindbladian} and \eqref{eq:Wjk}.
Equation~\eqref{eq:dotF} also contains the photon event operators
\begin{equation}
\label{eq:Jop}
{\cal J}_{l\alpha}^s=\sum_{jk}W_{jk}^{l\alpha s}{Y_{qjk}}
\end{equation}
which will be used to compute the current operator.

The mean currents ($n=1$) and higher order ($n>1$) correlations are obtained by averaging $n$-th derivatives of ${\cal F}(\chi)$ with respect to $i\chi_{l\alpha}$. Solving Eq.~\eqref{eq:dotF} may however be a hard task. Also, one is usually interested only in the first few cumulants. In those cases, we can obtain them recursively by performing a Taylor expansion:
\begin{equation}
{\cal F}=\rho+\sum_{l\alpha}i\chi_{l\alpha}{\cal F}_{l\alpha}^{(1)}-\frac{1}{2}\sum_{ll'\alpha\alpha'}\chi_{l\alpha}\chi_{l'\alpha'}{\cal F}_{l\alpha,l'\alpha'}^{(2)}+\dots
\label{eq:Fexp}
\end{equation}
up to the order of the desired correlation and introduce it
into Eq.~\eqref{eq:dotF}. Alternative approaches have been considered as well~\cite{christian,segal_current_2018}. As we are interested in the first two order moments:
\begin{align}
\label{eq:avNl}
\langle N_{l\alpha}\rangle&=\Tr{\cal F}_{l\alpha}^{(1)},\\
\label{eq:avNlNlp}
\langle N_{l\alpha}N_{l'\beta}\rangle&=\Tr{\cal F}_{l\alpha,l'\beta}^{(2)},
\end{align}
a second order expansion will be sufficient. 
With Eq.~\eqref{eq:Fexp}, we get a hierarchy of equations of motion. At zero-th order we recover the master equation for the reduced density matrix:
\begin{align}
\label{eq:F0}
\dot\rho&={\cal L}\rho.
\end{align}
The first and second order equations read
\begin{align}
\label{eq:F1}
\dot{\cal F}_{l\alpha}^{(1)}&={\cal I}_{l\alpha}^{}\rho+{\cal LF}_{l\alpha}^{(1)},
\end{align}
and
\begin{gather}
\label{eq:F2}
\begin{aligned}
\dot{\cal F}_{l\alpha,l'\beta}^{(2)}&={\cal I}_{l\alpha}^{}{\cal F}_{l'\beta}^{(1)}+{\cal I}_{l'\beta}^{}{\cal F}_{l\alpha}^{(1)}+\frac{1}{2}{\cal LF}_{l\alpha,l'\beta}^{(2)}\\
&+\delta_{ll'}\delta_{\alpha\beta}\left({\cal I}_{l\alpha}^{+}\rho+\frac{1}{2}{\cal LF}_{l\alpha,l'\beta}^{(2)}\right), 
\end{aligned}
\end{gather}
respectively. Using Eq.~\eqref{eq:Jop}, we define the current operator ${\cal I}_{l\alpha}\equiv{\cal I}_{l\alpha}^-$ from:
\begin{equation}
\label{eq:currOp}
{\cal I}_{l\alpha}^\pm\equiv{\cal J}_{l\alpha}^+\pm{\cal J}_{l\alpha}^-.
\end{equation}
With the solutions of Eqs.~\eqref{eq:F0}, \eqref{eq:F1} and \eqref{eq:F2} we get the system density matrix and the corresponding moments of the number of emitted particles, Eqs.~\eqref{eq:avNl} and \eqref{eq:avNlNlp}, whose time derivatives give the mean currents and their fluctuations, see Eqs.~\eqref{eq:IN} and \eqref{eq:corrN}. Note that we do not need to solve Eq.~\eqref{eq:F2} explicitly. Note also that, when replacing Eqs.~\eqref{eq:avNl} and \eqref{eq:avNlNlp} into Eq.~\eqref{eq:corrN}, the projection of ${\cal F}_{l\alpha}^{(1)}$ onto the kernel of ${\cal L}$ is canceled out when taking the trace~\cite{franz}. It is hence convenient to solve the equation for the perpendicular component: $\Upsilon_{l\alpha}={\cal F}_{l\alpha}^{(1)}-\rho\Tr{\cal F}_{l\alpha}^{(1)}$:
\begin{equation}
\dot\Upsilon_{l\alpha}=({\cal I}_{l\alpha}-I_{l\alpha})\rho+{\cal L}\Upsilon_{l\alpha},
\end{equation}
instead of \eqref{eq:F1}. In the stationary regime, we are left with the system of equations:
\begin{align}
\label{eq:statrhi}
{\cal L}\rho&=0\\
\label{eq:statF}
{\cal L}\Upsilon_{l\alpha}^{}&=-({\cal I}_{l\alpha}^{}-I_{l\alpha}^{})\rho,
\end{align}
which, completed by the conditions $\Tr\rho=1$ and $\Tr\Upsilon_{l\alpha}=0$, can be solved by simple linear algebra. Then, we can write the frequency resolved particle currents:
\begin{equation}
I_{l\alpha}=\Tr({\cal I}_{l\alpha}\rho)
\end{equation}
and the correlators:
\begin{equation}
S^N_{l\alpha,l'\beta}=\Tr\left[{\cal I}_{l\alpha}\Upsilon_{l'\beta}+{\cal I}_{l'\beta}\Upsilon_{l\alpha}+\delta_{ll'}\delta_{\alpha\beta}{\cal I}_{l\alpha}^+\rho\right].
\end{equation}
With this expression we can compute the correlations of particles with different frequencies in different detectors. The heat currents and the auto- ($l=l'$) and crosscorrelations ($l\neq l'$) are finally given by Eqs.~\eqref{eq:dotQ} and \eqref{eq:SQll}.
We can verify that the conservation of energy imposes
\begin{equation}
\label{eq:sumQS}
\sum_l\dot{Q}_l=0\quad{\rm and}\quad
\sum_{ll'}S_{ll'}=0,
\end{equation}
both for currents and fluctuations. 

\section{The demon as a nonequilibrium environment}
\label{sec:3teff}

Following the ideas of Refs.~\cite{ndemon,fatemeh}, it makes sense to compare the operation of the system with the efficiency of a three-bath configuration for which the demon terminals are replaced by a single reservoir, E, at temperature $T_E$. In that case, the simultaneous refrigeration and heat pumping of the two system reservoirs is not possible (in the absence of work sources). Even worse, under the demon conditions, here expressed as $\dot{Q}_E=0$, neither cooling nor pumping occurs. However, one can look at the efficiency of the separate operations when allowing for a finite $\dot{Q}_E$ (the case of a usual three-reservoir setup). 
For instance, the device can operate as an absorption refrigerator that extracts heat from a cold reservoir at temperature $T_C$ by putting it in contact with a hot reservoir at temperature $T_H$ and dumping heat into the environment E. 
To highlight the role of the warm demon, we define the efficiency in terms of the heat flowing into the environment (rather than the usual definition in terms of the heat from the hot bath) as
\beq
\tilde\eta_{abs,eq}=\frac{\dot{Q}_R}{-\dot{Q}_M}\leq\frac{T_C}{T_E}\frac{T_H-T_E}{T_H-T_C},
\eeq
where the inequality assumes that the three baths are in internal equilibrium. 
For the definition of $\tilde\eta_0$ in Eq.~\eqref{eq:relaxdem}, we replace $(T_H,T_C,T_E)$ by $(T_L,T_R,T)$.


\bibliography{biblio.bib}

\end{document}